\documentclass{ws-ijmpa}

\begin{document}

\markboth{Tomasz ~Ro{\.z}ek}
{Threshold Hyperon Production at COSY-11}

%
\catchline{}{}{}{}{}
%

\title{Threshold Hyperon Production at COSY-11
}

\author{\footnotesize T.~RO{\.Z}EK, D.~GRZONKA, K.~KILIAN, P.~KOWINA,
W.~OELERT, T.~SEFZICK, P.~WINTER, M.~WOLKE, P.~W{\"U}STNER}
\address{IKP and ZEL Forschungszentrum J{\"u}lich, D-52425 J{\"u}lich,
Germany} 

\author{\footnotesize  M.~SIEMASZKO, W.~ZIPPER}
\address{Institute of Physics, University of Silesia, Katowice 40-007, Poland}

\author{\footnotesize R.~CZY{\.Z}YKIEWICZ, M.~JANUSZ,
L.~JARCZYK, B.~KAMYS, P.~KLAJA, P.~MOSKAL, C.~PISKOR--IGNATOWICZ, J.~PRZERWA,
J.~SMYRSKI } \address{Nuclear Physics Department, Jagellonian
University, Cracow 30-059, Poland} 

\author{\footnotesize H.H.~ADAM, A.~KHOUKAZ, R.~SANTO, A.~T{\"A}SCHNER}
\address{IKP, Westf{\"a}che Wilhelms-Universit{\"a}t, D-48149
M{\"u}nster, Germany} 

\author{\footnotesize A.~BUDZANOWSKI} \address{Institute of Nuclear
Physics,  Cracow 31-342, Poland}

\maketitle

\pub{Received (Day Month Year)}{Revised (Day Month Year)}

\begin{abstract}

\vspace*{-0.3cm}

The $\Lambda$, $\Sigma^{0}$ and $\Sigma^{+}$ hyperon production in NN
collisions is studied at the \mbox{COSY -- 11} installation in order to
investigate the production mechanism as well as to extract information
about the Y-N interaction. 

\vspace*{-0.1cm}

\end{abstract}

\vspace*{0.4cm}

COSY-11 is an internal magnetic spectrometer experiment at the COoler
SYnchroton and storage ring COSY in J\"ulich. It is equipped with
scintillator hodoscopes and drift chambers for charged particle
detection~\cite{bra96} and a scintillator/lead sandwich detector for neutrons~\cite{neutr}.

\begin{figure}
\centerline{\psfig{file=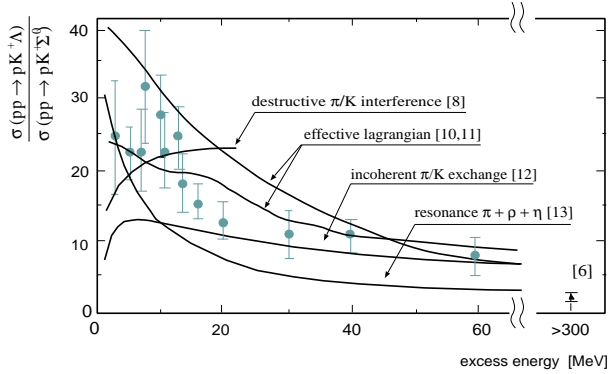,width=0.65\textwidth}}
\vspace*{8pt}
\caption{\label{all_models}Cross section ratio for $\Sigma^0$ and
$\Lambda$ production in the threshold region. The data are compared to
different models descriptions.}
\end{figure}

The $\Sigma^0$ and $\Lambda$ hyperon production near the kinematical
threshold was studied by the COSY-11 collaboration in $pp\rightarrow
pK^+\Lambda/\Sigma^0$ reactions. Data points, 16 for the $\Lambda$
and 13 for the $\Sigma^0$ channel, were taken in the excess energy range
between 0.68 MeV and 59.3 MeV for the $\Lambda$ hyperon and between 2.8
MeV and 59.1 MeV for $\Sigma^0$~\cite{bal98,sew99,kow02}. 
The cross section ratio $\sigma (pp\rightarrow pK^+\Lambda ) /\sigma
(pp\rightarrow pK^+\Sigma^0 )$ below excess energies of 15~MeV
was measured to be around 28 in contrast to the value of
about 2.5 determined for excess energies higher than $Q = 300$ MeV
\cite{bald88}. The ratio for higher energies is in good agreement with
the $\Lambda/\Sigma^0$ isospin relation, which is 3 (see~figure~\ref{all_models}). 

To explain this unexpected threshold behaviour, various theoretical
scenarios within meson exchange models were proposed. Calculations
have been performed with pion and kaon exchange added coherently
\cite{gas99,gas01} or incoherently~\cite{sib99}, including the
excitation of nucleon resonances~\cite{shy01,sib00,shy04} and heavy meson
exchange ($\rho$, $\omega$ and $K^*$)~\cite{shy01,shy04}. Although the
various descriptions differ even in the dominant basic reaction
mechanism, all more or less reproduce the trend of an increase in the
cross section ratio in the threshold region. The present data are not
sufficient to definitely exclude possible explanations. Further
studies e.g. for the other isospin projections will help to understand
the threshold hyperon production. To be more specific lets consider
the J\"ulich meson exchange model\cite{gas99,gas01} where calculations
are available for other $\Sigma$ channels. Within this model the large
cross
section ratio $\sigma (\Lambda) / \sigma (\Sigma ^0)$ at threshold is reproduced by
a destructive interference of $\pi$ and $K$ exchange
amplitudes. Calculations of the $\Sigma^+$ production in this model
predict a factor of three higher cross section compared to the
$\Sigma^0$ channel for a destructive and a factor of three lower for a
constructive interference, a clear experimentally accessible signal. The
ratio between $\Sigma^+$ and $\Sigma^0$ production will also differ
strongly if the dominant production mechanism runs via an intermediate
$N^*$ excitation or not.\\

\begin{figure}
\centerline{\psfig{file=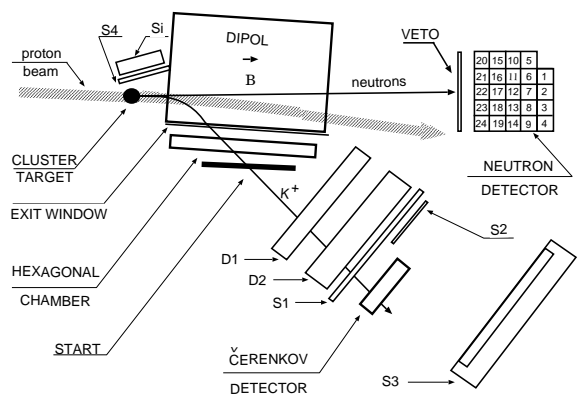,width=0.6\textwidth}}
\vspace*{8pt}
\caption{\label{COSY} COSY--11 detection setup with the
superimposed tracks of kaon and neutron from the $pp\rightarrow nK^+\Sigma^+$ reaction.}
\end{figure}

Recently the $\Sigma^+$ production was
measured at the COSY-11 installation via $pp\rightarrow nK^+\Sigma^+$
at Q = 13 MeV and Q = 60 MeV \cite{pac02}. The $\Sigma^+$ is
identified via the missing mass technique, by detecting the remaining
reaction products - $K^+$ and neutron. The experimental resolution of
the missing mass determination depends on the reconstruction accuracy
of the four-momentum vectors for the registered neutrons and
kaons. The momentum vector of the $K^+$ meson can be established by
tracking the $K^+$ trajectory reconstructed from signals registered in
the drift chambers (D1 and D2 in the figure~\ref{COSY}) through the
magnetic field back to the target point. Assuming a hit in the neutron
detector being due to a neutron, the four-momentum vector of the neutron is
given by the measured velocity, the direction (given by the first
hit module) of the neutron which can be reconstructed with an
accuracy of at least the size of the module (9x9 $cm^2$) and the known
mass. The background from charged particles hitting the neutron
detector is discriminated by veto scintillators (VETO in
figure~\ref{COSY}).\\
From the Monte Carlo studies the identification of the $\Sigma^+$
events is expected to be comparable to the $\Sigma^0$ or $\Lambda$
channels\cite{pac02}. The data are presently under analysis.   

\vspace*{-0.1cm}
\section*{Acknowledgments}

The work has been supported by the European Community - Access to
Research Infrastructure action of the Improving Human Potential
Programme, by the DAAD Exchange Programme (PPP-Polen), by the Polish
State Committe for Scientific Research (grants No. 2P03B07123 and
PB1060/P03/2004/26) and by the Research Centre J{\"u}lich.

\end{document}